\documentclass{Interspeech}

\interspeechcameraready

\title{A Speech Production Model for Radar: Connecting Speech Acoustics with Radar-Measured Vibrations
}

\author[affiliation={1}]{Isabella}{Lenz}
\author[affiliation={1}]{Yu}{Rong}
\author[affiliation={1}]{Daniel}{Bliss}
\author[affiliation={2}]{Julie}{Liss}
\author[affiliation={1,2}]{Visar}{Berisha}

\affiliation{Electrical Engineering}{Arizona State University}{Tempe, USA}
\affiliation{College of Health Solutions}{Arizona State University}{Tempe, USA}
\email{ilenz@asu.edu}

\usepackage{comment}

\begin{document}

\maketitle

\begin{abstract}
    
Millimeter Wave (mmWave) radar has emerged as a promising modality for speech sensing, offering advantages over traditional microphones. Prior works have demonstrated that radar captures motion signals related to vocal vibrations, but there is a gap in the understanding of the analytical connection between radar-measured vibrations and acoustic speech signals. We establish a mathematical framework linking radar-captured neck vibrations to speech acoustics. We derive an analytical relationship between neck surface displacements and speech. We use data from 66 human participants, and statistical spectral distance analysis to empirically assess the model. Our results show that the radar-measured signal aligns more closely with our model filtered vibration signal derived from speech than with raw speech itself. These findings provide a foundation for improved radar-based speech processing for applications in speech enhancement, coding, surveillance, and authentication.
\end{abstract}

\section{Introduction}

Alternative sensors and sensor fusion present an opportunity to extend and augment microphones capabilities, unlocking new possibilities in speech processing. Non-acoustic systems such as bone conducting microphones, accelerometers and vibrometers \cite{dekens2013body,bone2004,McCree2007Fusion}, electromyography sensors \cite{meltzner2018development}, and electromagnetic sensors \cite{Ma2019AuditoryRadar,holzrichter1998speech, sami2020spying} have been used to capture signals related to speech such as vocal fold, and vocal tract articulator motions, muscle contraction and sound-wave induced vibrations. These signals have been utilized in speech enhancement \cite{li201394,zhang2020viblive,demiroglu2004noise}, coding \cite{Quatieri2006Nonacoustic, holzrichter1998speech}, silent-speech recognition \cite{digehsara2022user}, surveillance \cite{Ma2019AuditoryRadar} and authentication  \cite{liu2023wavoid,xu2019waveear} applications. 

Millimeter Wave (mmWave) radar has recently been proposed as a promising supplement to acoustic sensing, addressing key limitations of traditional microphone-based systems. Unlike microphones, which degrade in noisy environments, mmWave radar is immune to ambient acoustic noise. When combined with microphones, it enhances speech processing by providing spatial context and capturing biomechanical vibrations useful for speech enhancement. Compared to other non-acoustic sensors like bone-conducting microphones and accelerometers, mmWave radar offers a key advantage: it does not require physical contact, making it less intrusive. Additionally, mmWave radar outperforms lower-frequency electromagnetic sensors due to its shorter wavelengths and wider fractional bandwidths, which enable higher spatial and temporal resolution.

To fully leverage mmWave radar as an acoustic sensor, it is essential to have an  understanding of the physiological signals being measured by the sensor. There is consensus in the literature that radar captures a motion signal related to vocal vibration \cite{Geiger2018160Flex,xu2019waveear,liu2023wavoid,Ma2019AuditoryRadar}. There is, however, limited analytical and empirical study on how the radar captured vibration signal relates to the acoustic signal. 

\begin{figure}
\centering
\includegraphics[width = .65\linewidth]{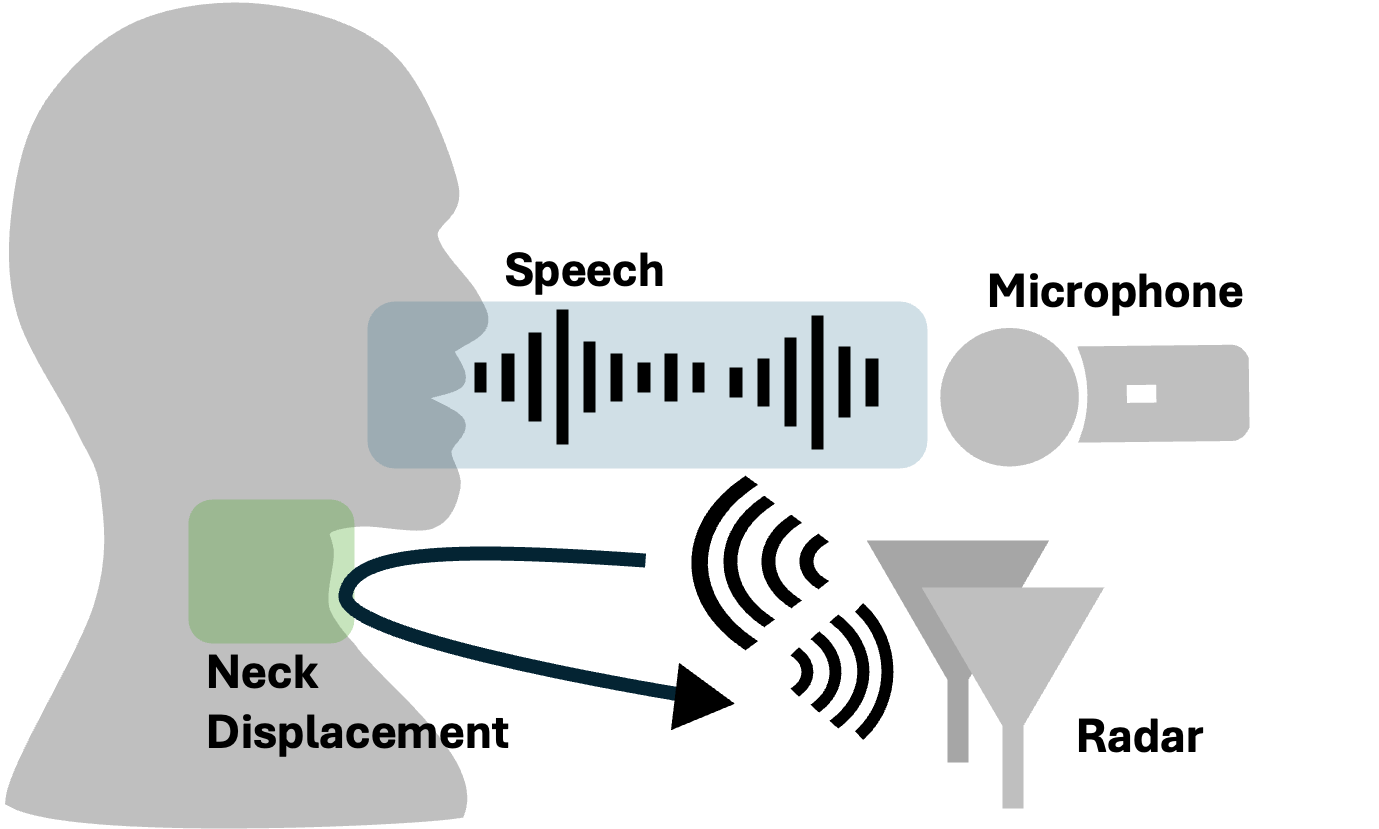}
\caption{The radar pulses mmWaves and receives the reflection from neck displacement to recover the vibration signal. The microphone captures the speech pressure wave that has been shaped by the vocal tract. \label{introcartoon}}
\end{figure}

Speech production involves a complex series  of physiological motions including respiration, vocal fold vibrations, and articulatory shaping. The airflow from the lungs drives vocal fold vibrations, while the articulators—such as the lips, tongue, and jaw—shape this sound into intelligible speech. During this process, the vocal folds induce vibrations on the surface of the neck. The vibrating surface of the neck, during speech production introduces a micro-doppler shift in a transmitted mmWave. A radar recovers this vibration signal by analyzing the micro-doppler of reflected mmWaves. The reflection of the mmWaves from neck displacement and the pressure wave produced by speech are depicted in Fig. \ref{introcartoon}.


We present a mathematical derivation that relates neck vibrations (displacement) measured by a mmWave radar system to the acoustic signal recorded by the microphone. We connect the source-filter speech production model with a radar model and demonstrate how the speech can be processed to arrive at acoustic correlate of the radar signal. We empirically asses our model using a dataset collected from 66 human participants. Statistical spectral distance analysis confirms that the radar-measured vibrations are more closely related to the model transformed vibration signal than to the raw speech signal. 

\begin{figure*}[ht]
\centering
\includegraphics[width = .85\textwidth]{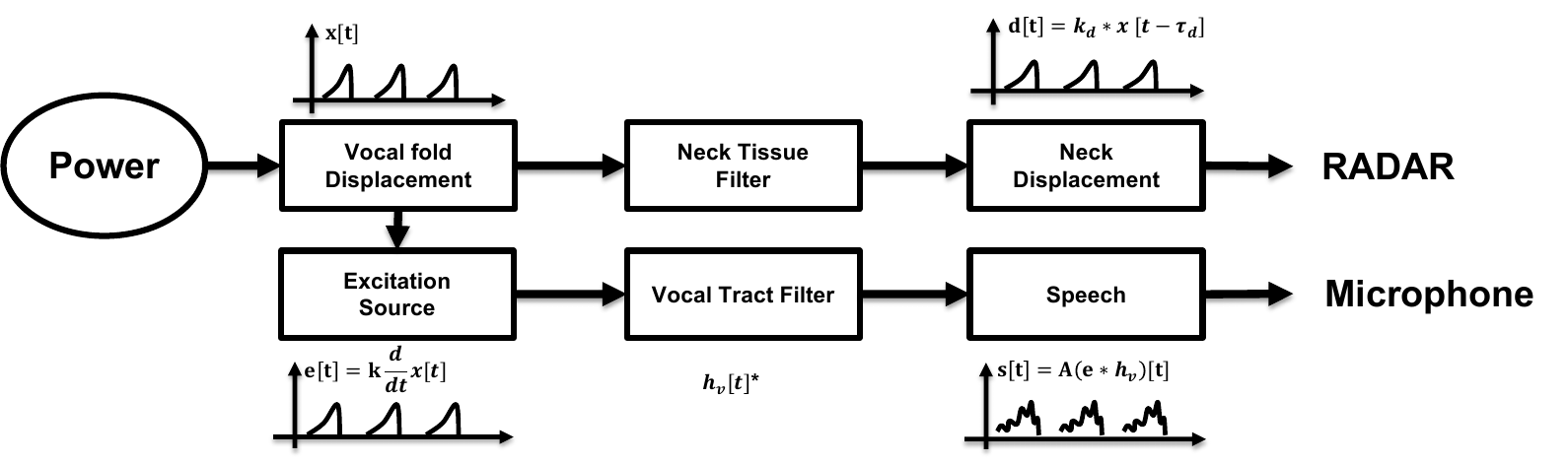}
\caption{Block diagram summarizing the derived model. Neck vibration $d(t)$ and raw speech $s(t)$ are functions of the vocal fold displacement. \label{modelbd}}
\end{figure*}

This model provides a foundation for improved radar-based speech processing and fusion with traditional microphones. By connecting the radar data with a speech-correlated representation, it enables more interpretable radar representations and more effective fusion between radar-sensed speech and microphone-sensed speech. Our work addresses the lack of prior radar-based speech models, serving to improve the integration of radar in downstream speech processing tasks (e.g. speech enhancement, silent speech interfaces, etc.)

\section{Methods}

This subsection presents a mathematical model relating the neck vibration signal during voiced speech captured by a radar sensor to the raw speech signal measured by a microphone. In the case of a human speaker, these two signals originate from the same source, vocal displacement $x(t)$, but travel through different channels. A model of the channels is therefore required to recover the source signals from the two modalities for comparison.

\subsection{Assumptions} Radar can reliably measure vibrations only during voiced speech, as vocal fold oscillations induce surface vibrations on the neck. During unvoiced speech, however, the vocal folds remain open and do not generate periodic vibrations, making radar-based measurements less reliable.

As is common in speech processing, we assume a stationary model for voiced speech, meaning that within a short analysis frame, voiced sounds can be approximated as sustained phonation. This allows us to assume that the pressure differential above and below the vocal folds remains constant within each frame, simplifying the modeling of both the radar and acoustic signal pathways.

\begin{figure*}
\centering
\includegraphics[width = .9\linewidth]{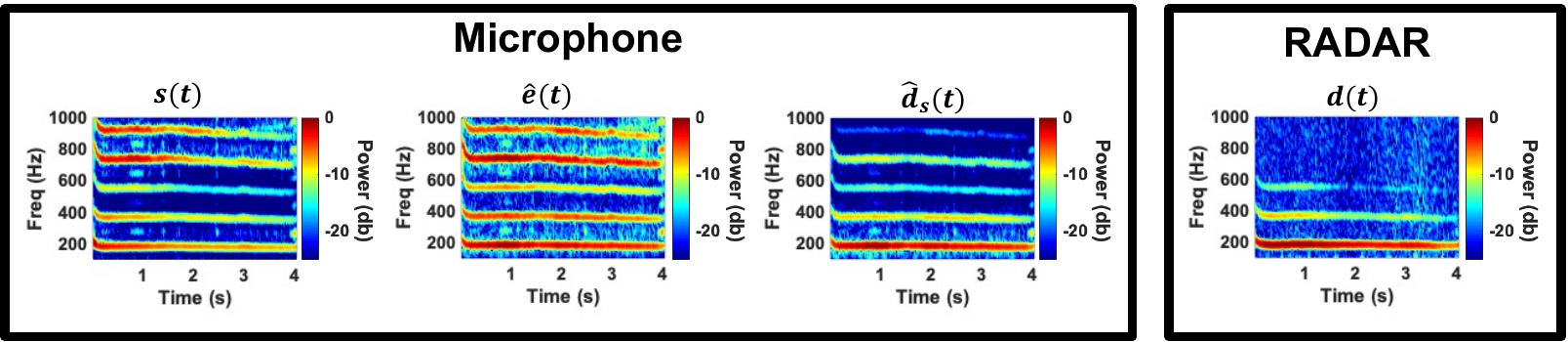}
\caption{ Comparison of time aligned sustained phonation signals. The left box shows the spectrograms from the microphone raw speech signal $s(t)$ , the estimated the excitation $\hat{e}(t)$ and the model filtered neck displacement $\hat{d}(t)$. The right box shows the spectrograms from the radar neck vibration signals. \label{spectcomp}}
\end{figure*}

\subsection{Neck Vibration from Vocal Displacement}
During voiced sound production, the neck vibrates because the vocal folds open and close
at the rate of the fundamental frequency. This opening and closing causes neck displacements (vibrations). The displacement of the neck surface $d(t)$ is related to the vocal fold displacement $x(t)$ as
    \begin{align}
        d(t) = k_d *x(t-\tau_d) \label{eq:ndx},
    \end{align}
    where \( k_d \) is a proportionality constant accounting for tissue transmission characteristics.
    and \(\tau_d \) is the time delay due to the propagation through the
    neck tissues.

\subsection{Speech From Vocal Vibration}
 During voiced sound production, the raw speech signal $s(t)$  is a function of vocal fold displacement $x(t)$ and a vocal tract filter $h(t)$
 
 The airflow through the glottis, is related to the vocal fold displacement $x(t)$ by a scaling constant $K$. The scaling constant combines a proportionality constant relating displacement to area and a pres- sure differential constant. The speech excitation signal $e(t)$ represents the changes in glottal airflow over time,
    \begin{align}
      e(t) &= K\frac{d}{dt}x(t).\label{eq:exciteog}
    \end{align}
 
In source-filter modeling, we typically assume the raw speech signal $s(t)$ is modeled as the
convolution of the excitation signal $e(t)$ and the vocal tract’s impulse response $h(t)$ as 
    \begin{align}
        s(t) &= (e*h)(t).\label{eq:speechexcite}
    \end{align}
\subsection{Connecting Raw Speech and Neck Displacement}\label{sec:modelconnect}

In the previous subsections, we have established models for both the raw speech signal and the radar-measured neck vibrations during sustained phonation as functions of vocal fold displacement $x(t)$. In this section we present the explicit steps to recover the model filtered neck vibration signal $\hat{d}$ from the  raw speech signal. This is summarized in Algorithm block \ref{connection}. We first estimate frame-based vocal tract impulse response ($h$) using standard LPC analysis described in \cite{jackson2013digital}. We use the coefficients ($h$) to inverse filter $s(t)$ and estimate excitation signal $\hat{e}(t)$ as
    \begin{align}
      \hat{e}(t) = (s*h^{-1})(t). \label{eq:xspeech}
    \end{align}

We then estimate the  vocal fold displacement taking the integral of the excitation signal
    \begin{align}
      \hat{x}(t) = \int \hat{e}(t) dt. 
    \end{align}   

We cross correlate vocal fold displacement ($ \hat{x}(t)$) and radar neck vibration ($d(t)$) to estimate the propagation delay $\tau_d$. We then delay and scale $\hat{x}(t)$ to recover the model filtered neck displacement as
    \begin{align}
      \hat{d}(t) &=
      \frac{\hat{x}(t-\tau_d)}{K}  .
      \label{fig:dttospeech}
    \end{align}

    \begin{algorithm}
  \caption{Connecting Raw Speech and Neck Displacement}\label{connection}
  \begin{algorithmic}[1]
    \Require Raw Speech $s[t]$, radar $d[t]$ window length $N$, step size $\Delta$
    \State $L \gets \text{length}(s)$
    \State $N_w \gets \lfloor (L - N)/\Delta \rfloor + 1$ 
    \For{$i = 0$ to $N_w - 1$}
    \State $n = [i\Delta : i\Delta + N - 1]$
          \State ${h}\gets s[n]$ \Comment{impulse response}
          \State $\hat{e}[n]\gets s[n],{h}$ \Comment{inverse filter}
          \State $\hat{x}[n]\gets \hat{e}[n]$ \Comment{integration}
          \State ${\tau}_d \gets \hat{x}[n] * d[n]$ \Comment{cross correlation}
          \State $\hat{d}[n] t\gets \hat{x}[n],{\tau}_d$ \Comment{model filtered output}
    \EndFor\label{windows}
    \State \textbf{return} $\hat{d},\hat{e}$
  \end{algorithmic}
\end{algorithm}

\begin{figure}[b]
\centering
\includegraphics[width = .85\linewidth,trim={1cm 1.6cm 0 .6cm},clip]{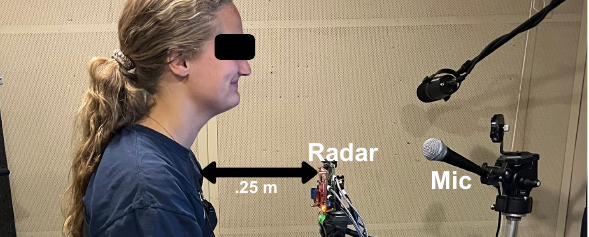}
\caption{Example experimental set-up. Data is collected in a sound booth. A subject is seated at .$25$m the radar, the radar is approximately aligned with the subjects collarbone. A microphone is set near ($<25$cm) the radar.\label{expsetup}}
\end{figure}
\subsection{Hardware Systems} 
In our study we use an off the shelf microphone (PGA48-LC) and an off the shelf mmWave FMCW MIMO radar sensor (Texas Instruments AWR1843 Radio Frequency (RF) Array and DCA1000EVM Data Capture Adapter). We configure the FMCW chirp signal to have a 77 GHz starting frequency  and 3.6 GHz bandwidth. The time of each chirp is 60 microseconds and the pulse repetition frequency is 2kHz.

\subsection{Signal processing implementation}
We first recover neck displacement signal $d(t)$ using standard radar signal processing chain described in \cite{lenz2023contactless}. We in parallel collect the raw speech signal $s(t)$ with a microphone. We sample both signals at 2kHz. The sampling rate is chosen for two reasons: radar hardware limitation, and human physiology (men's vocal folds can vibrate from 90 - 500 Hz and Women's vocal folds can vibrate from 150 -1000 Hz \cite{voice}). We then estimate the excitation $\hat{e}(t)$ and the model filtered neck displacement $\hat{d}(t)$ using the raw speech signal using the following steps in Algorithm block \ref{connection}.


\section{Results}

Equation \ref{fig:dttospeech} establishes a relationship between the displacement signal recovered by radar and an inverse-filtered, integrated version of the speech signal. To evaluate the accuracy of this model, we compare the radar signal against three other signals:

\begin{itemize}
    \item \textbf{Raw speech signal}: direct microphone recording.
    \item \textbf{Inverse-filtered excitation signal}: isolating the source excitation.
    \item \textbf{Model-derived acoustic signal}: described in Section \ref{sec:modelconnect}.
\end{itemize}

Comparisons are conducted on a frame-by-frame basis after aligning the radar and acoustic signals in time. The evaluation metric is the \textit{log-spectral distance (LSD)}, which quantifies the spectral similarity between signals. Our hypothesis is that the \textbf{model-derived acoustic signal will exhibit the lowest LSD} when compared to the radar signal, indicating that it provides the closest match to the physiological vibrations measured by radar.

\subsection{Experimental Setup and Dataset}
To perform empirical assessment, we collect data from 66 subjects. The study is approved by ASU IRB , with each participant providing consent and recieving a \$10 gift card for their participation. Previous respiratory or speech disorder diagnosis disqualifies participation. Participant demographic data was not collected. Data is collected in a sound booth.  The participant is seated $0.25$ meters from the radar, which is approximately aligned with the subjects collarbone. The microphone is placed near ($<25$cm) the radar. The participant is instructed to sustain a phonation of the vowel sound $\backslash$\textipa{A:}$\backslash$ for approximately 5 seconds. They are then prompted to read a series of 8 sentences. The experimental set up is depicted inFig.\ref{expsetup}.

\begin{figure}(t)
\centering
\includegraphics[width = .9\linewidth]{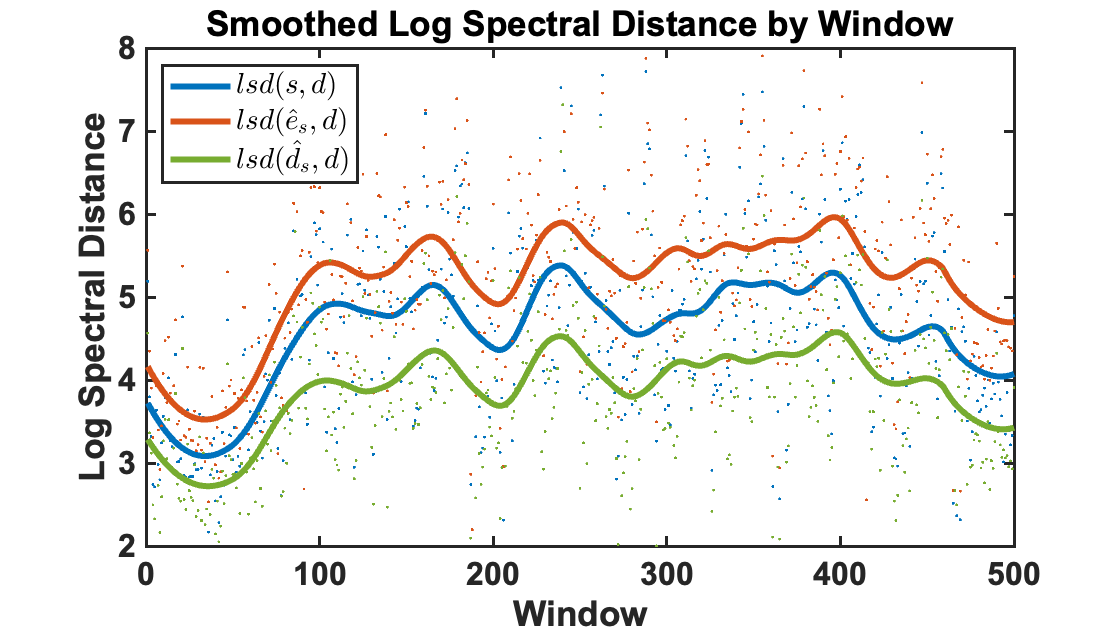}
\caption{Loess regression smoothed log spectral distance from the radar signal $d(t)$ over windows of speech for one subject. The blue line represents the  raw Speech signal $s(t)$, the orange line represents the estimated exercitation signal $\hat{e}(t)$ and the green line represents the model filtered signal $\hat{d}(t)$\label{loess}}.
\end{figure}

\subsection{Statistical Analysis}
The derived model is based on knowledge of human physiology and speech production and the physics of radio frequency propagation. We aim to use empirical data to validate the accuracy and identify limitations of the model, derived in section \ref{sec:modelconnect}

We perform paired sample $t$-tests to evaluate the similarity between the radar- captured neck displacement $d(t)$ and: (a) the raw speech signal $s(t)$; (b) an estimated excitation signal $\hat{e}(t)$; (c) and the model-filtered neck vibration $\hat{d}(t)$. The null hypothesis is that the mean difference between any two sets of observations is zero. This implies that applying the derived model transformation to the speech signal does not increase spectral similarity to the radar-captured neck displacement.

To accomplish this we need a method to assess the spectral similarity of between two time dependent signals. We first window $s(t)$ , $\hat{e}(t)$, $\hat{d}(t)$ and $d(t)$ into $25$ms windows with $50\%$ overlap and calculate the normalized power spectrum of each signal in each window.

\begin{figure}[b]
\centering
\includegraphics[width = .9\linewidth]{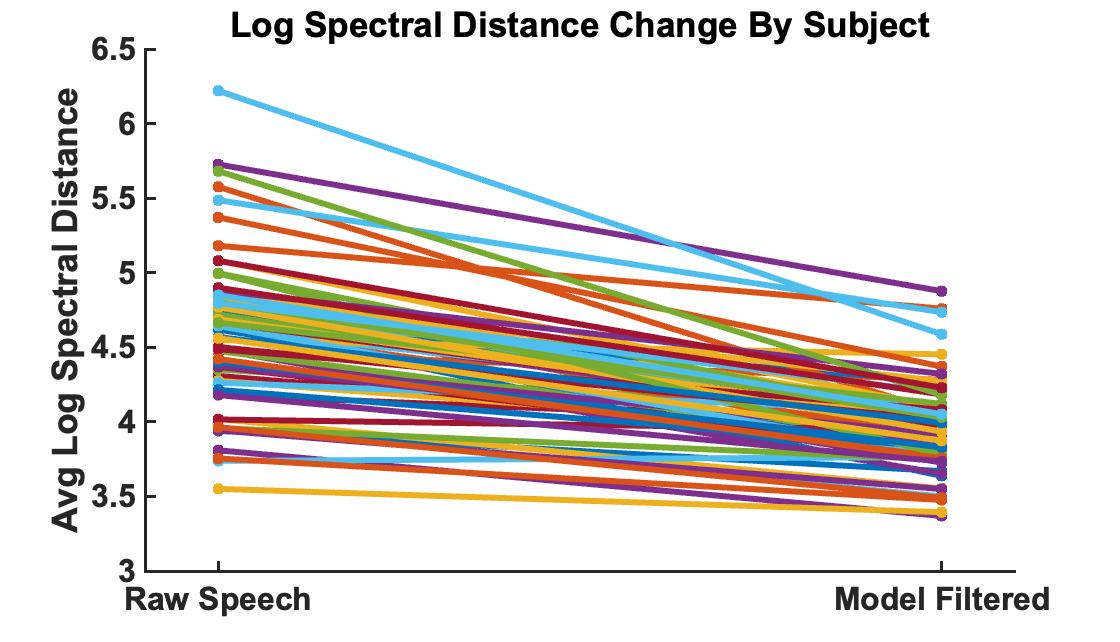}
\caption{The average log spectral distance (LSD) between the raw Speech signal $s(t)$ and the radar signal $d(t)$ over all windows of voiced speech for each subject is plotted as the first point. The average LSD between the model filtered signal $\hat{d}(t)$ and the radar signal $d(t)$ is plotted as the second point.\label{LSDchange}}.
\end{figure}

We then calculate the mean log spectral distance (LSD) between the normalized power spectra of $s(t)$ $\hat{e}(t)$, $\hat{d}(t)$ and the normalized power spectra of $d(t)$ in each window n. Mean LSD is a widely used speech comparison metric because (a.) spectral comparison is more robust to low-level noise and small time shifts in the waveform that significantly impact the time-domain Euclidean distance and (b.) the log transformation reduces the impact of small variations in low-energy regions, making LSD more noise-resistant \cite{gray1976distance}.

We repeat the signal transformation and mean LSD calculation for each subject in the data set, giving us 66 pairs of observation of the mean LSD between the speech signal and neck displacement, the estimated excitation signal and neck displacement and the model filtered neck displacement and neck displacement. 

We perform a paired sample t-test on the observations  (a) the mean LSD between the speech signal and neck displacement and (b) the model filtered neck displacement and and neck displacement.  We perform a second  paired sample t-test  on the observations of (a) the mean LSD between the estimated excitation signal and neck displacement and (b) the model filtered neck displacement and and neck displacement. The null hypothesis is that the mean difference between the two sets of observations is zero. Rejecting the null hypothesis implies the model filtered neck displacement is more spectrally comparable to the radar neck vibration signal than the raw speech signal and estimate excitation signal respectively.

\subsection{Statistical Analysis Results}


The descriptive statistics for the  LSDs for each of the three comparisons can be found in \ref{table:descstat}. The mean of the mean LSD is greatest for estimated excitation signal and lowest for the model filtered neck displacement.

\begin{table}[h]
  \caption{Descriptive Statistics\label{table:descstat}}
  \setlength{\tabcolsep}{0.7\tabcolsep}
  \centering
  \begin{tabular}{ *{5}{l} }
    \toprule
    & \textbf{Mean} & \textbf{N} & \textbf{STD}& \textbf{SEM}\\
    \midrule
    Raw Speech & 4.5923 &  66  &  0.4964 & 0.0593\\
    Estimated Excitation & 5.3441 & 66 & 0.3961 & 0.0473 \\
    Model Filtered & 3.9740 &  66  & 0.3054 & 0.0365\\
    \bottomrule
  \end{tabular}
\end{table}

Fig. \ref{loess} shows the Loess regression smoothed LSD from the radar signal $d(t)$ over windows of speech for one subject. The model filtered neck vibration signal (green line) has lower LSD than the estimated excitation (orange) and the raw speech (blue). Spectrogram examples of the signals at each state in the transformation for one subjects prolonged 'ahh' can be seen in figure. \ref{spectcomp}.

The results of the two paired $t$-tests are shown in Table \ref{table:descstat}. Both $t$-tests reject the null hypothesis with greater that 99.99$\%$ confidence, implying the model filtered neck displacement is more spectrally comparable to the radar neck vibration signal than the raw speech signal and estimated excitation.  
\begin{table}[h]
  \caption{Paired Samples Correlation \label{table:corr}}
  \setlength{\tabcolsep}{0.7\tabcolsep}
  \centering
  \begin{tabular}{ p{3cm}llll}
    \toprule
    & \textbf{$t$-stat} & \textbf{N} & \textbf{STD} & \textbf{P-Value}\\
    \midrule
    Raw Speech v. Model Filtered & -15.626 & 66   & 0.324 & $<0.001$\\
    Estimated Excitation v. Model Filtered & -54.065 & 66   & 0.707 & $<0.001$\\
    \bottomrule
  \end{tabular}
\end{table}

As this was a paired-samples $t$-test, we can evaluate the within-subject differences for each evaluation.Fig.\ref{LSDchange} plots, for each subject, the mean LSD for the raw speech  as the first point and the mean LSD for the model filtered neck vibration as the second point. Negative slope implies greater spectral similarity when the model is applied. 

The totality of these results demonstrate that there is an increase in the spectral similarity between the model-filtered acoustic signal and the neck vibration signal measured by radar.

\section{Conclusion}
In this work, we introduced a mathematical framework that relates radar-measured neck vibrations to speech signals. 
Empirical assessment, using data collected from human subjects, confirms that radar-based measurements more closely aligns with the estimated neck displacement than with both the raw speech signal and the estimated excitation signal.  Our work enables advancement in radar-based speech processing, with applications in speech enhancement, coding, surveillance, and authentication.

\bibliographystyle{IEEEtran}
\bibliography{mybib}

\begin{thebibliography}{10}
\providecommand{\url}[1]{#1}
\csname url@samestyle\endcsname
\providecommand{\newblock}{\relax}
\providecommand{\bibinfo}[2]{#2}
\providecommand{\BIBentrySTDinterwordspacing}{\spaceskip=0pt\relax}
\providecommand{\BIBentryALTinterwordstretchfactor}{4}
\providecommand{\BIBentryALTinterwordspacing}{\spaceskip=\fontdimen2\font plus
\BIBentryALTinterwordstretchfactor\fontdimen3\font minus \fontdimen4\font\relax}
\providecommand{\BIBforeignlanguage}[2]{{%
\expandafter\ifx\csname l@#1\endcsname\relax
\typeout{** WARNING: IEEEtran.bst: No hyphenation pattern has been}%
\typeout{** loaded for the language `#1'. Using the pattern for}%
\typeout{** the default language instead.}%
\else
\language=\csname l@#1\endcsname
\fi
#2}}
\providecommand{\BIBdecl}{\relax}
\BIBdecl

\bibitem{dekens2013body}
T.~Dekens and W.~Verhelst, ``Body conducted speech enhancement by equalization and signal fusion,'' \emph{IEEE transactions on audio, speech, and language processing}, vol.~21, no.~12, pp. 2481--2492, 2013.

\bibitem{bone2004}
Z.~Zhang, Z.~Liu, M.~Sinclair, A.~Acero, L.~Deng, J.~Droppo, X.~Huang, and Y.~Zheng, ``Multi-sensory microphones for robust speech detection, enhancement and recognition,'' in \emph{2004 IEEE International Conference on Acoustics, Speech, and Signal Processing}, vol.~3, 2004, pp. iii--781.

\bibitem{McCree2007Fusion}
A.~McCree, K.~Brady, and T.~F. Quatieri, ``Multisensor dynamic waveform fusion,'' in \emph{2007 IEEE International Conference on Acoustics, Speech and Signal Processing - ICASSP '07}, vol.~4, 2007, pp. IV--577--IV--580.

\bibitem{meltzner2018development}
G.~S. Meltzner, J.~T. Heaton, Y.~Deng, G.~De~Luca, S.~H. Roy, and J.~C. Kline, ``Development of semg sensors and algorithms for silent speech recognition,'' \emph{Journal of neural engineering}, vol.~15, no.~4, p. 046031, 2018.

\bibitem{Ma2019AuditoryRadar}
Y.~Ma, H.~Hong, H.~Zhao, H.~Li, C.~Gu, and X.~Zhu, ``Speech recovery based on auditory radar and webcam,'' in \emph{2019 IEEE MTT-S International Microwave Biomedical Conference (IMBioC)}, vol.~1, 2019, pp. 1--3.

\bibitem{holzrichter1998speech}
J.~Holzrichter, G.~Burnett, L.~Ng, and W.~Lea, ``Speech articulator measurements using low power em-wave sensors,'' \emph{The Journal of the Acoustical Society of America}, vol. 103, no.~1, pp. 622--625, 1998.

\bibitem{sami2020spying}
S.~Sami, Y.~Dai, S.~R.~X. Tan, N.~Roy, and J.~Han, ``Spying with your robot vacuum cleaner: eavesdropping via lidar sensors,'' in \emph{Proceedings of the 18th Conference on Embedded Networked Sensor Systems}, 2020, pp. 354--367.

\bibitem{li201394}
S.~Li, Y.~Tian, G.~Lu, Y.~Zhang, H.~Lv, X.~Yu, H.~Xue, H.~Zhang, J.~Wang, and X.~Jing, ``A 94-ghz millimeter-wave sensor for speech signal acquisition,'' \emph{Sensors}, vol.~13, no.~11, pp. 14\,248--14\,260, 2013.

\bibitem{zhang2020viblive}
L.~Zhang, S.~Tan, Z.~Wang, Y.~Ren, Z.~Wang, and J.~Yang, ``Viblive: A continuous liveness detection for secure voice user interface in iot environment,'' in \emph{Proceedings of the 36th Annual Computer Security Applications Conference}, 2020, pp. 884--896.

\bibitem{demiroglu2004noise}
C.~Demiroglu and D.~V. Anderson, ``Noise robust digit recognition using a glottal radar sensor for voicing detection.'' in \emph{INTERSPEECH}, 2004, pp. 813--816.

\bibitem{Quatieri2006Nonacoustic}
T.~Quatieri, K.~Brady, D.~Messing, J.~Campbell, W.~Campbell, M.~Brandstein, C.~Weinstein, J.~Tardelli, and P.~Gatewood, ``A 94-ghz millimeter-wave sensor for speech signal acquisition,'' \emph{IEEE Transactions on Audio, Speech, and Language Processing}, vol.~14, no.~2, pp. 533--544, 2006.

\bibitem{digehsara2022user}
P.~A. Digehsara, J.~V.~P. de~Menezes, C.~Wagner, M.~B{\"a}rhold, P.~Schaffer, D.~Plettemeier, and P.~Birkholz, ``A user-friendly headset for radar-based silent speech recognition.'' in \emph{INTERSPEECH}, 2022, pp. 4835--4839.

\bibitem{liu2023wavoid}
T.~Liu, F.~Lin, C.~Wang, C.~Xu, X.~Zhang, Z.~Li, W.~Xu, M.-C. Huang, and K.~Ren, ``Wavoid: Robust and secure multi-modal user identification via mmwave-voice mechanism,'' in \emph{Proceedings of the 36th Annual ACM Symposium on User Interface Software and Technology}, 2023, pp. 1--15.

\bibitem{xu2019waveear}
C.~Xu, Z.~Li, H.~Zhang, A.~S. Rathore, H.~Li, C.~Song, K.~Wang, and W.~Xu, ``Waveear: Exploring a mmwave-based noise-resistant speech sensing for voice-user interface,'' in \emph{Proceedings of the 17th Annual International Conference on Mobile Systems, Applications, and Services}, 2019, pp. 14--26.

\bibitem{Geiger2018160Flex}
M.~Geiger, D.~Schlotthauer, and C.~Waldschmidt, ``Improved throat vibration sensing with a flexible 160-ghz radar through harmonic generation,'' in \emph{2018 IEEE/MTT-S International Microwave Symposium - IMS}, 2018, pp. 123--126.

\bibitem{jackson2013digital}
L.~B. Jackson, \emph{Digital Filters and Signal Processing: With MATLAB{\textregistered} Exercises}.\hskip 1em plus 0.5em minus 0.4em\relax Springer Science \& Business Media, 2013.

\bibitem{lenz2023contactless}
I.~Lenz, Y.~Rong, and D.~Bliss, ``Contactless stethoscope enabled by radar technology,'' \emph{Bioengineering}, vol.~10, no.~2, p. 169, 2023.

\bibitem{voice}
\BIBentryALTinterwordspacing
``Your voice and how it works.'' [Online]. Available: \url{https://med.umn.edu/ent/patient-care/lions-voice-clinic/about-the-voice/how-it-works}
\BIBentrySTDinterwordspacing

\bibitem{gray1976distance}
A.~Gray and J.~Markel, ``Distance measures for speech processing,'' \emph{IEEE Transactions on Acoustics, Speech, and Signal Processing}, vol.~24, no.~5, pp. 380--391, 1976.

\end{thebibliography}

\end{document}